\documentstyle[aps,preprint]{revtex}
\def\lsim{\mathrel{\raise2pt\hbox to 8pt{\raise -5pt\hbox{$\sim$}\hss{$<$}}}}

\newcommand{\ba}{\begin{eqnarray}}
\newcommand{\ea}{\end{eqnarray}}
\input BoxedEPS
\SetepsfEPSFSpecial
\begin{document}

\title{A Relativistic Symmetry in Nuclei} \author{
{Joseph N. Ginocchio}} \address {Theoretical Division, Los Alamos National Laboratory, Los Alamos, New Mexico 87545, USA}
\maketitle

\begin{abstract}
We review the status of a quasi - degenerate doublets in nuclei, called pseudospin doublets, which were discovered about thirty years ago
and the origins of which have remained a mystery, until recently.  We show that pseudospin doublets originate from an SU(2) symmetry of
the Dirac Hamiltonian which occurs when the scalar and vector potentials are opposite in sign but equal in magnitude. Furthermore, we
survey the evidence that pseudospin symmetry is approximately conserved for a Dirac Hamiltonian with realistic scalar and vector
potentials. We briefly discuss the relationship of pseudospin symmetry with chiral symmetry and the implications of pseudospin symmetry
for the antinucleon spectrum in nuclei. 

\end{abstract}


\maketitle

\section{Introduction}

Dick Slansky and I met almost thirty years ago as assistant professors at Yale University. Although,for the most part, Dick's published
research at that time was in hadron collisions \cite {dick}, he became interested in group theory and symmetry, particularly applied to
elementary particles physics, during his stay at Yale. I had similar interest in group theory and symmetry, but primarily applied to the
many - body problem, particularly in nuclear physics. In the mid seventies Dick and I each accepted positions at Los Alamos National
Laboratory and we continued to discuss symmetry in physics. At Los Alamos he published a number of papers in group theory including his
very popular review on ``Group Theory for Unified Model Building''
\cite {dick2}, and  also published a book on affine Lie algebras \cite {dick3}. In 1989 he became Theoretical Division Leader at the Los
Alamos National Laboratory, and, technically, my ``boss''. 

At the time we were at Yale I had heard about an observation that certain single -
nucleon levels in spherical nuclei were found to clump into quasi - degenerate doublets \cite {kth,aa}. The term pseudospin doublets was
applied to these states, although the quasi - degeneracies we assumed to be accidental. About two years ago I rexamined these pseudospin
doublets, the origin of which remained a mystery, and I discovered that they are a consequence of a relativistic symmetry \cite
{gino,ami}.  I discussed this revelation with Dick and he suggested that this symmetry may have some connection with chiral symmetry.  In
this paper I would like to discuss the progress that has been made in understanding pseudospin symmetry, including an enticing connection
with chiral symmetry via QCD sum rules.

\section{Pseudospin Symmetry}
The spherical shell model orbitals that were observed to be quasi - degenerate have non -
relativistic quantum numbers ($n_r$,
$\ell$, $j =
\ell + 1/2)$ and ($n_{r}-1, \ell + 2$, $j = \ell + 3/2$) where $n_r$, $\ell$, and $j$ are 
the single-nucleon radial, orbital, and total 
angular momentum quantum numbers, respectively
\cite {kth,aa}.  This doublet structure is expressed in
terms of a ``pseudo'' orbital angular momentum 
$\tilde{\ell}$ = $\ell$ + 1, the average of the orbital angular momentum of the two 
states in doublet and ``pseudo'' spin, $\tilde s$ = 1/2.
For example,
$(n_r s_{1/2},(n_r-1) d_{3/2})$ will have
$\tilde{\ell}= 1$ , $(n_r p_{3/2},(n_r-1) f_{5/2})$ will have $\tilde{\ell}= 2$, etc. 
These doublets are almost degenerate with
respect to pseudospin, since $j = \tilde{\ell}\ \pm \tilde s$ for the two states 
in the doublet; examples are shown in Figure 1. 
Pseudospin ``symmetry'' was shown
to exist in deformed nuclei as well \cite {bohr,draayer3} and has been used to explain
features of deformed nuclei, including superdeformation \cite {dudek} and identical bands
\cite {twin,stephens}. However, the origin of pseudospin symmetry remained a mystery and ``no deeper understanding of the origin of
these (approximate) degeneracies'' existed
\cite {ben}. A few years ago it was shown that relativistic mean field theories gave 
approximately the correct spin orbit splitting to
produce the pseudospin doublets \cite {draayer}. In this paper we shall review more recent
developements that show that pseudospin  symmetry is a relativistic symmetry \cite
{gino,gino2,ami}. 

\section{Symmetries of the Dirac Hamiltonian}

The success of the shell model implies that nucleons move in a mean field 
produced by the interactions between the nucleons.
Normally, it suffices to use the Schrodinger equation to describe the motion of the 
nucleons in this mean field.  However, in order to
understand the origin of pseudospin symmetry, we need to take into account the 
motion of the nucleons in a relativistic mean field and
thus use the Dirac equation. The Dirac Hamiltonian, H, with an external scalar, $V_S$, and 
vector, $V_V$, potentials is given by:
\begin{equation}
H =  {\vec{\alpha}} \cdot {\vec p} + \beta (m + V_S) + V_V ,
\label {dirac}
\end{equation}
where we have set 
%${h \over 2\pi}
$\hbar = c =1$,  ${\vec \alpha}$, $\beta $ are the usual Dirac matrices
\cite {mul}, $m$ is the nucleon mass, and $\vec p$ is the three momentum. The Dirac Hamiltonian is invariant under an SU(2) algebra for two
limits: 
 $V_S - V_V$ = constant and $V_S + V_V$ = constant. For finite nuclei the constant will be zero since each potential
will go to zero at large distances \cite{bell}. (For infinite nuclear matter the constant could be non-zero.) 

The generators for the SU(2) algebra, ${\hat {S}}_i$, which commute with the
Dirac Hamiltonian, $[\,H\,,\, {\hat { S}}_i\,] = 0$, for the case when $V_S = V_V$ are given by 
\cite{bell}
\begin{equation}
{\hat {S}}_i = {{\vec \alpha\cdot \vec p \ {\hat s}_i \ \vec \alpha\cdot \vec p} \over {p^2}} \
{(1 -\beta)\over 2} + {\hat  s}_i\, {(1 + \beta)\over 2}, 
\label {sbell}
\end{equation}
where ${\hat s}_i = \sigma_i/2$ are the usual spin generators and $\sigma_i$ the Pauli matrices. This reduces to 
\ba
{\hat {S}}_i = \left ( {{\hat {s}_i} \atop 0 } { 0 \atop { {\hat{\tilde s}}_i}}\right ).
\label {sgen}
\ea
where
\ba
{\hat {\tilde s}}_i} = U_p\ {\hat {s}_i \ U_p = {2\,{\vec s \cdot \vec p} \over p^2}\ p_i -
{\hat {s}}_i . 
\label {ugen}
\ea
In (\ref {ugen}) $U_p = \, {{ \vec \sigma\cdot \vec p} \over p}$
is the momentum-helicity unitary operator introduced in \cite {draayer}
that accomplishes the transformation from the normal shell model space to the pseudo shell model space while preserving
rotational, parity, time - reversal, and translational invariance. This symmetry limit leads to spin doublets, but we know that the
spin orbit splitting in nuclei is very large so this limit is not applicable to nuclei. However we shall see that it is relevant for the
antinucleon spectrum.

We shall show that, in the limit of $V_S = -V_V$, the conserved symmetry is pseudospin symmetry. The generators for the SU(2) algebra,
${\hat {\tilde S}}_i$, which commute with the Dirac Hamiltonian, $[\,H\,,\, {\hat {\tilde S}}_i\,] = 0$, for the case when $V_S = -V_V$
are given by 
\cite{bell}
\begin{equation}
{\hat {\tilde S}}_i = {{\vec \alpha\cdot \vec p \ {\hat s}_i \ \vec \alpha\cdot \vec p} \over {p^2}} \
{(1 +\beta)\over 2} + {\hat  s}_i\, {(1 - \beta)\over 2}. 
\label {bell}
\end{equation}
This reduces to 
\ba
{\hat {\tilde S}}_i = \left ( {{\hat {\tilde s}_i} \atop 0 } { 0 \atop { {\hat { s}}_i}}\right ).
\label {gen}
\ea

For $V_S = - V_V$, the eigenfunctions of the Dirac Hamiltonian, 
$H \Psi_{\tau,\tilde\mu} = {\cal E}_{\tau} \Psi_{\tau,\tilde\mu}$ are
doublets (${\tilde S} =1/2$,
$\tilde\mu =\pm 1/2$) with respect to the SU(2) generators ${\hat {\tilde S}}_i$ of Eq. (\ref{gen})
\ba
{\hat {\tilde S}}_z\,\Psi_{\tau,\tilde \mu} &=& \tilde\mu\, \Psi_{\tau,\tilde\mu} ~, \nonumber\\
{\hat {\tilde S}}_{\pm}\,\Psi_{\tau,\tilde \mu} &=& {\sqrt{(1/2 \mp {{\tilde \mu}})( 3/2 \pm {{\tilde \mu}}) }}\, \Psi_{{\tau, \tilde \mu}
\pm 1} ~,
\qquad\qquad (\,V_S = - V_V\,)
\label {ugen2}
\ea
where ${\hat {\tilde S}}_{\pm} = {\hat {\tilde S}}_{x} \pm i{\hat {\tilde S}}_y$. The eigenvalue $\tau$ refer to the other necessary
quantum numbers.

The fact that the the pseudo - spin generators (\ref {gen}) have only the spin operator ${\hat {s}}_i$ operating on the lower component
of the Dirac wave function has the consequence that the spatial wavefunctions for the two states in the pseudospin doublet are
identical in the limit of $V_S = - V_V$ to within an overall phase. 

This symmetry for $V_S = -
V_V$ is general and applies to deformed nuclei as well as spherical
nuclei. In the case for which the potentials are spherically symmetric, the Dirac Hamiltonian
has an additional invariant SU(2) algebra; namely, the pseudo - orbital angular momentum,
\ba
{\hat {\tilde L}}_i =
\left ( {\hat {\tilde \ell_i} \atop 0 }
{ 0 \atop { {\hat \ell}_i} }\right ),
\label {jgen}
\ea
where ${\hat {\tilde \ell}}_i = U_p\, {\hat \ell}_i$ $U_p$, ${\hat\ell}_i = \vec r \times 
\vec p$.
In this limit, the Dirac wave functions are eigenfunctions of the
Casimir operator of this algebra,
${ {\hat {\tilde L}}\cdot {\hat {\tilde L}}}\, |\Psi_{\tau,{\tilde \ell},j,m_j}\rangle
= {\tilde\ell} ({\tilde \ell} + 1) |\Psi_{\tau,{\tilde \ell},j,m_j}\rangle $, where we have used a 
coupled basis,
and $j$ is the eigenvalue of the total
angular momentum operator ${\hat J}_i = {\hat {\tilde L}}_i + {\hat {\tilde S}}_i,\ {{\hat
J}\cdot{\hat J}}\, |\Psi_{\tau, {\tilde \ell},j,m_j}\rangle = j(j +
1)|\Psi_{\tau, {\tilde\ell},j,m_j}\rangle $, and $m_j$ is the eigenvalue of ${\hat J}_z$. Thus
pseudo-orbital angular momentum as well as pseudospin are conserved in the spherical
limit and $V_S = -V_V$. 
From (\ref{jgen}), we see that the lower component wave function 
will have spherical harmonic of rank ${\tilde \ell}$ coupled to spin to give total angular 
momentum $j$. Since ${\vec \sigma\cdot \vec p}$
conserves the total angular momentum but $\vec p$ changes the orbital angular 
momentum by one unit because of parity conservation, 
the upper component also has total angular momentum $j$, 
but orbital angular momentum $\ell = {\tilde \ell} \pm
1$. If $j = {\tilde \ell} + 1/2$, then it follows that $\ell = {\tilde \ell} + 1$, whereas if $j = {\tilde \ell} - 1/2$, then $\ell = {\tilde \ell} -
1$. This agrees with the pseudospin doublets originally observed \cite{kth,aa} and discussed at 
the beginning of this paper.  

For axially symmetric deformed nuclei, there is
a U(1) generator corresponding to the pseudo-orbital angular momentum
projection along the symmetry axis which is conserved in addition to the
pseudospin for $V_S = - V_V$,
\begin{equation}
{\hat{\tilde \lambda}} = \left ( {{\hat {\tilde \Lambda}} \atop 0 } { 0 \atop {{\hat \Lambda}} }
\right ),
\label {Lgen}
\end{equation}
where ${\hat {\tilde \Lambda}} = U_p\ \hat \Lambda\ U_p$. In this case the Dirac wave functions are eigenfunctions of ${\hat
{\tilde \lambda}}$,
${\hat
{\tilde \lambda}}\ |\Psi_{\tau,{\tilde \Lambda},{ \Omega}}\rangle = {\tilde \Lambda}
|\Psi_{\tau,{\tilde \Lambda},\Omega}\rangle $, where ${\Omega}$ is total angular momentum
projection, ${ \Omega} = {\tilde \Lambda} + {\tilde \mu}$, which has the same value for the
upper and lower components since $\vec\sigma\cdot \vec p$ conserves the total
angular momentum projection. Thus
${\Omega} = {\tilde \Lambda} \pm 1/2$, corresponding exactly to the quantum
numbers of the pseudospin doublets
for axially deformed nuclei discussed in \cite {bohr,draayer3}.

However, the exact symmetry limit can not be realized in nuclei, because, if $V_S = - V_V$, 
there are no Dirac bound valence states and hence nuclei
can not exist.  However, we now show that pseudospin symmetry is approximately realized 
for $V_S \approx - V_V$.

\section{Free Nucleons}

For a free Dirac nucleon, $V_S = V_V = 0$, and hence both symmetries are valid. The free nucleon
wavefunction is 

\ba
\Psi_{p,\mu} = {\cal N} \left ( {u_{\mu} \atop {{2 \ \mu\  p_z \ u_{\mu}} \over {(m +E_p)}}} \right ) e^{i(p_zz-E_pt\ )}
\label {hel}
\ea
where $u_{\mu}$ are the Dirac spinors $u_{1/2} = ({1 \atop 0}), u_{-1/2} = ({0 \atop 1})$, $E_p = \sqrt{p^2 + m^2}$, and ${\cal N}$ is the
normalization. This wavefunction has definite helicity and hence is an eigenfunction of
the helicity operator,
\ba
U_p\Psi_{p, \mu} &=& 2 \mu\, \Psi_{p,\mu}. 
\label {hel}
\ea
The wavefunctions with $\mu = \pm 1/2$ are doublets for ${\hat S}_i$ and, since ${\hat {\tilde S}}_i = U_p {\hat { S}}_i U_p$, the
wavefunctions
${\tilde
\Psi}_{p, {\tilde \mu}}= U_p\Psi_{p,\mu} = 2 \mu\, \Psi_{p,\mu}$ are doublets for ${\hat {\tilde S}}_i$, the pseudospin. 

\section{Realistic mean fields}

A near equality in the magnitude
of mean fields, $V_S \approx - V_V$, is a universal feature of the relativistic mean field 
approximation (RMA) of 
relativistic field theories with interacting nucleons and mesons \cite {wal} and relativistic 
theories with nucleons interacting via zero range
interactions 
\cite{mad}, as well as a consequence of QCD sum rules \cite{furn}. We shall discuss QCD sum rules in the next section. 

Recently realistic
relativistic mean fields were shown to exhibit approximate pseudospin symmetry in both 
the energy spectra and wave functions \cite {gino2,ring,arima}. 
In Fig. 2 we show the energy splittings 
between pseudospin doublets normalised by $2{{\tilde \ell}} + 1$ as a function of
the average binding energy $\langle \epsilon \rangle = (\epsilon_{{\tilde \ell} + 1/2} +
\epsilon_{{\tilde \ell} - 1/2})/2$. 
We see that the energy splitting for the same pseudo - orbital angular momentum decreases as
the radial quantum increases; that is, as the binding energy decreases. Also for the same
binding energy, the energy splitting increases as the pseudo - orbital angular 
momentum increases. These features follow from the square well potential \cite {gino}. In Table 1 we
tabulate some of the calculated energy splittings compared to the measured splittings. What is
interesting is that the measured splittings are smaller than those calculated by
relativistic mean field theory indicating that pseudospin symmetry breaking is overestimated
by the relativistic mean field approximation.

Pseudospin doublets will manifest themselves for deformed potentials as well when $V_S \approx - V_V$. In Figure 3 the single
particle (s.p.) energies of the doublets are plotted versus the deformation. We see that for each value of pseudo asymptotic quantum
$[{\tilde N}{\tilde n_3}{\tilde \Lambda} ]$ numbers there is a quasi-degenerate pseudospin doublet with $\Omega = {\tilde {\Lambda}} \pm
1/2$.

As mentioned in the last section the relativistic SU(2) pseudo-spin symmetry implies that the
spatial wavefunction for the lower component of the Dirac wavefunctions will be equal in shape
and magnitude for the two states in the doublet. For spherical nuclei the Dirac
wavefunction for the two states in the doublet are $\Psi_{\tau,j = \tilde {\ell} + 1/2 , m} =
(g_{\tau, \tilde {\ell} + 1/2} [Y_{\tilde {\ell} + 1}\chi]_m^{j = \tilde {\ell} + 1/2 },
if_{\tau,\tilde {\ell} + 1/2} [Y_{\tilde {\ell}}\chi]_m^{j = \tilde {\ell} + 1/2 }),
\ \Psi_{ \tau,j = \tilde {\ell} - 1/2, m} = (g_{\tau, \tilde {\ell} - 1/2} [Y_{\tilde {\ell} - 1}
\chi]_m^{(j = \tilde {\ell} - 1/2)}, if_{\tau, \tilde {\ell} - 1/2}[Y_{{\tilde
{\ell}}}\chi]_m^{(j =
\tilde {\ell} - 1/2)})$ where $g, f$ are the radial wave functions,
$Y_{\ell}$ are the spherical harmonics, $\chi$ is a two-component Pauli spinor, and
$[\dots]^{(j)}$ means coupled to angular momentum $j$. For a square well potential, the
overall phase between the two amplitudes will be a minus sign \cite {gino} so we expect that, in the
symmetry limit for realistic potentials, $f_{\tau,\tilde {\ell} + 1/2}(r) = - f_{\tau,\tilde {\ell} -
1/2}(r)$.  In Figure 4  we see that, for realistic zero range potentials, $f_{ \tau,{\tilde {\ell}} + 1/2}(r) \approx - f_{\tau,{\tilde
{\ell} - 1/2}}(r)$  \cite {gino2}.

These results are also valid for the relativistic mean field approximation to a nuclear field theory with meson exchanges \cite {ring}. 
In Figure 5a) the pseudospin doublets in the vicinity of the Fermi surface for neutrons and protons are shown. The upper (g) and lower
components (f) of the pseudospin doublets are also shown in Figure 5b,c,d.  While the upper components are very different with different
nodal stucture, the lower components are almost identical.  We also note that the lower components are small with respect to the
upper components, which is consistent with the non-relativistic shell model.

\section{ QCD Sum Rules}

Applying QCD sum rules in nuclear matter, the scalar and vector self-energies were determined to be \cite {furn}

$$
\Sigma_s = - {{4{\pi}^2 \sigma_N\rho_{N}} \over {m^2} m_q}
$$
\begin{equation}
\Sigma_v = {{32{\pi}^2 \rho_{N}} \over {m^2}}, 
\label {qcd} 
\end{equation}
where $\rho_{N}$ is the nuclear density, and $m_q$ the quark mass.  $\sigma_N$ is the sigma term which arises
from the breaking of chiral symmetry \cite {cheng}.  The ratio then becomes

\begin{equation}
{\Sigma_s \over \Sigma_v} = - {\sigma_N \over 8 m_q}. 
\label {ratio} 
\end{equation}
For reasonable values of $\sigma_N $ and quark masses, this ratio is close to -1. The implication of these results is that
chiral symmetry breaking is responsible for the scalar field being approximately equal in magnitude to the vector field, thereby producing
pseudospin symmetry.

\section{Antinucleon Spectrum}

The antinucleon states are obtained by charge conjugation, $C$, applied to the negative energy eigenstates of the Dirac Hamiltonian \cite
{mul}.  This leads to a spectrum which has quasi - degenerate spin doublets, not pseudospin doublets. This follows from the
fact that, under charge conjugation,
\ba
C^{\dagger}{\hat {\tilde S}}_i, C = \left ( {{\hat{ s}_i} \atop 0 } { 0 \atop { {\hat{\tilde s}}_i}}\right ) = {\hat { S}}_i.
\label {cgen}
\ea
Thus in (\ref {cgen}) the spin operator ${\hat{ s}_i}$ operates on the upper component and hence the spatial wavefunctions for the upper
components of the states in spin doublet will be very similar. Likewise for spherical nuclei, the pseudo-orbital angular momentum
goes into the orbital angular momentum, and for axially deformed nuclei, pseudo-orbital projection goes into orbital projection along the
body fixed z axis. 

This symmetry in the antinucleon spectrum also follows from the fact that the antinucleon potentials are $\bar {V}_S = C^{\dagger}V_S\ C
=V_S$, and
$ {\bar V}_V =  C^{\dagger}V_V\ C = - V_V$.  Thus  ${\bar V}_S \approx {\bar V}_V $ and the symmetry of the Dirac Hamiltonian generated by
(\ref {sgen}) applies and spin doublets are produced in the antinucleon spectrum \cite {bell}.

\section{Summary}

We have shown that pseudospin symmetry is a broken SU(2) symmetry of the Dirac Hamilonian which describes the motion of nucleons in 
realistic scalar and vector mean field potentials, $V_S \approx - V_V$.  This symmetry predicts that the spatial wavefunctions of the
lower components for states in the doublet will be very similar in shape and size and this has been substantiated by relativistic mean
field approximations of relativistic nuclear field theories and relativistic nuclear Lagrangians with zero range interactions. This
symmetry has been linked via QCD sum rules to chiral symmetry breaking in nuclei. Finally, the antinucleon spectrum is shown
to have a spin symmetry rather than a pseudospin symmetry.

Future applications of pseudospin symmetry will involve the testing of the wavefunction through the relationships between transitions
within pseudospin doublets which follow from pseudospin symmetry\cite {gino3}.

\section{Acknowlegements}

This work was supported by the
United States Department of Energy under contract
W-7405-ENG-36.\\
 
%\noindent {\bf References}\\

\begin{table}
\caption{ $^{208}Pb$ pseudospin doublet energy splittings}
\vspace{18pt}
\vbox {\tabskip 2em plus 3em minus 1em \halign to \hsize{\hfil  
#\hfill && #\hfil \hfil 
\cr ${\tilde \ell}$ & $ps \  doublets $ & $\epsilon_{\tilde \ell + 1/2}-\epsilon_{\tilde \ell - 1/2}$ (RMA){\cite {gino2}}
&$\epsilon_{\tilde \ell + 1/2}-\epsilon_{\tilde \ell - 1/2}$ (EXP) &
\cr
\noalign{\hrule}
\noalign {\vskip 12pt}\ & neutrons &  &\  
\cr 4 & $0h_{9/2} - 1f_{7/2}$ & 2.575 & \ 1.073 
\cr 2 & $1f_{5/2}-2p_{3/2}$ &  0.697 &  -0.328 
&\cr
\noalign{\hrule}
\noalign {\vskip 12pt}\ & protons &  & \ 
\cr 3& $0g_{9/2} - 1d_{5/2}$ & \ 4.333 & \ 1.791 
\cr 1 & $1d_{3/2} - 2s_{1/2}$ & \ 1.247 & \ 0.351 &
\cr 
\noalign {\vskip 12pt} }}
\end{table}
\pagebreak

\begin{figure}
\caption{Examples of pseudospin doublets in the $^{208}Pb$ region. $n_r$ is the radial quantum number of the state, $\ell$ is the orbital
angular momentum, $j$ the total angular momentum.}
\label{1}
\end{figure}
\begin{figure}
\caption{Pseudospin doublet energy splittings normalised by $2{\tilde{\ell}} + 1$ as a function of the average 
binding energy $\langle \epsilon \rangle$.  $n_r$ is the radial quantum number of the state with the lower orbital quantum
number}
\label{2}
\end{figure}

\begin{figure}
\caption{Single - particle (s.p.) energies for neutron pseudospin partners as a function of deformation}
\label{3}
\end{figure}

\begin{figure}
\caption{$^{208}Pb$ lower component wavefunctions $f_{\tau,{\tilde \ell}- 1/2}$(r) (dash line), -$f_{\tau,{\tilde \ell} + 1/2}$(r) (dot
- dash line) for the
$(2s_{1/2},1d_{3/2})$  pseudospin doublet as a function of the radius r} 
\label{4}
\end{figure}
\begin{figure}
\caption{$^{208}Pb$: a) energy spectrum; b,c,d) upper (g) and lower (f) components of the Dirac wavefunction
for pseudospin doublets as a function of the radius r} 
\label{5}
\end{figure}
\HideDisplacementBoxes
\hskip1.0truein 
\BoxedEPSF{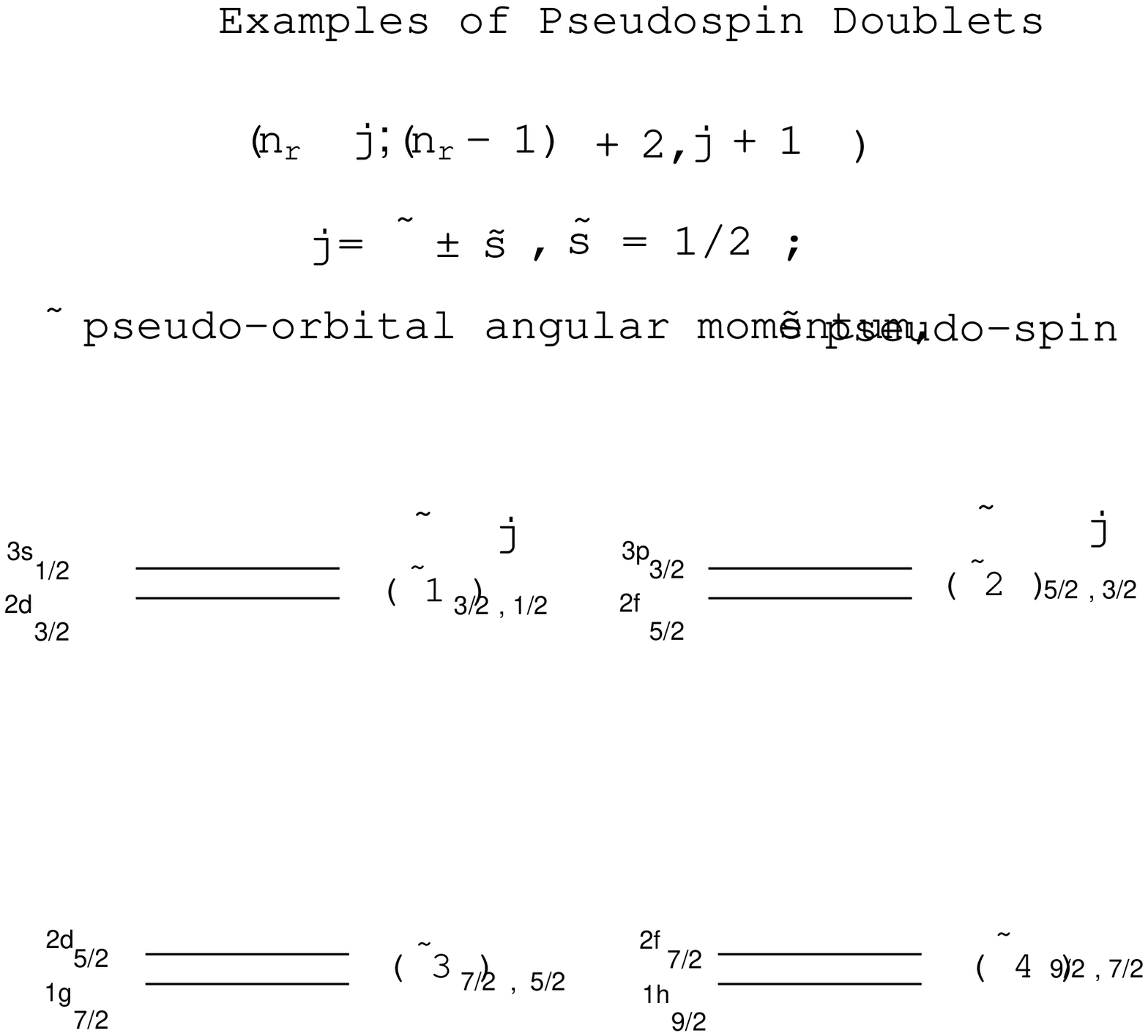 scaled 800}
\HideDisplacementBoxes
\hskip1.0truein 
\BoxedEPSF{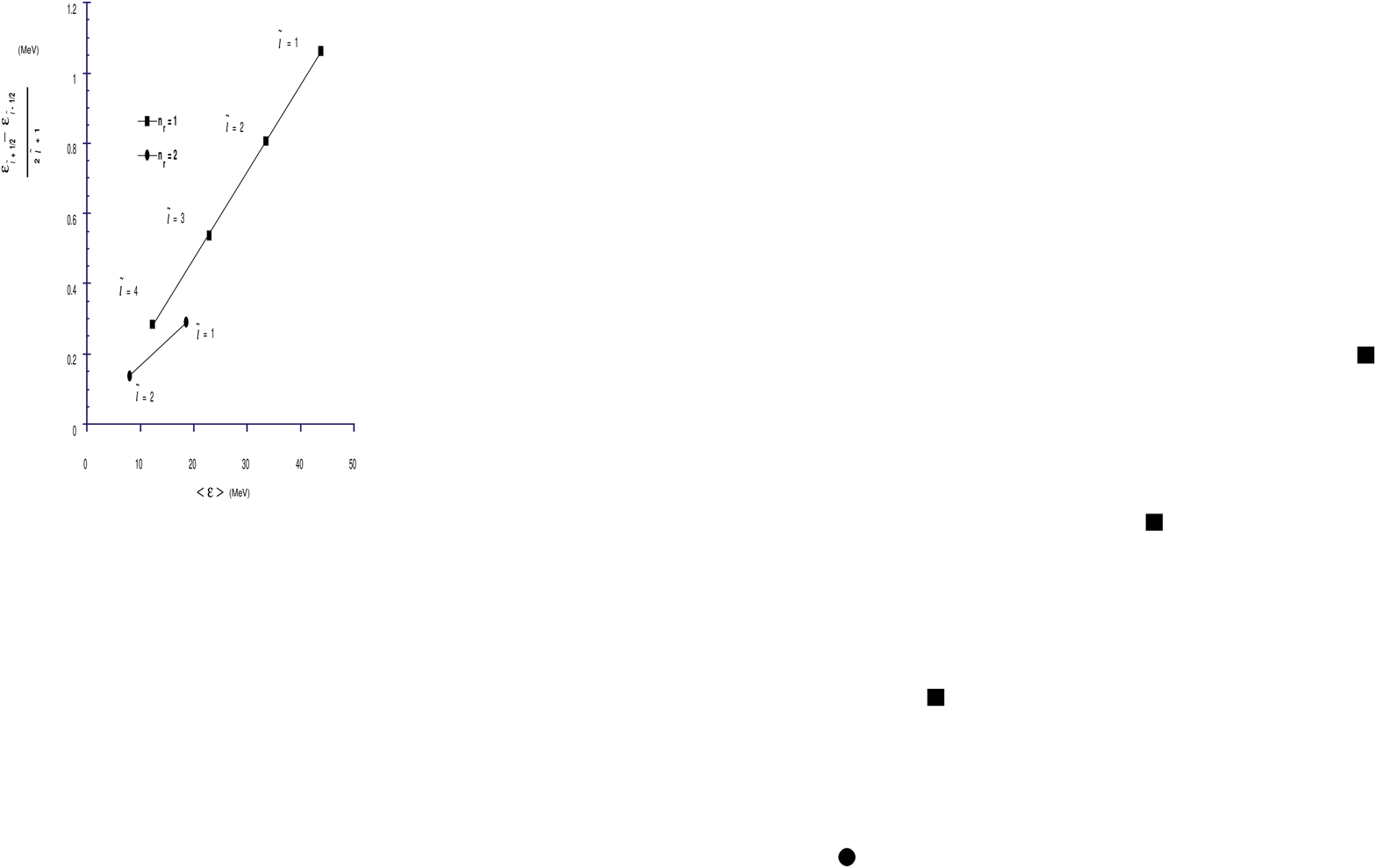 scaled 800}
\HideDisplacementBoxes
\hskip1.2truein 
\BoxedEPSF{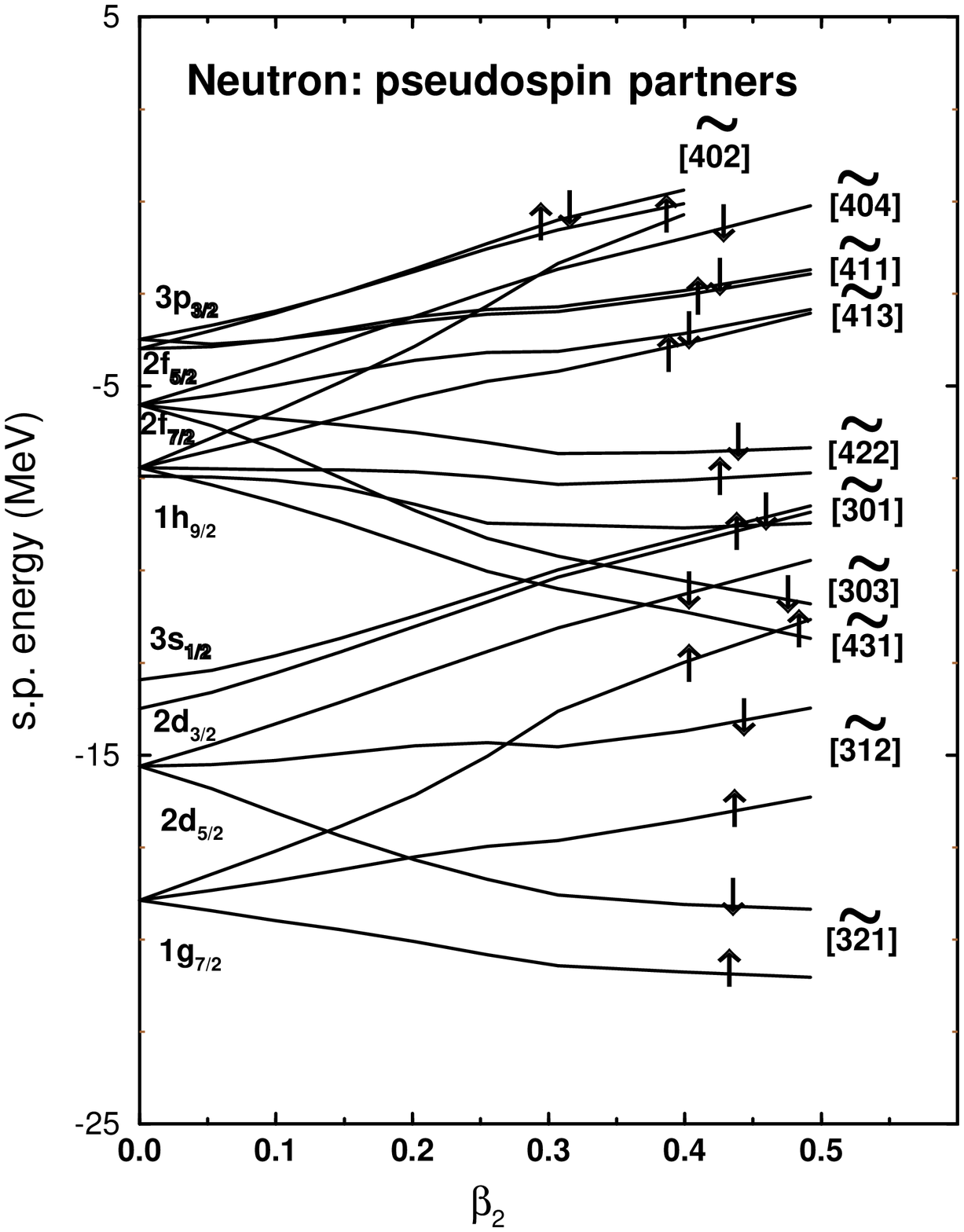 scaled 800}
\HideDisplacementBoxes
\hskip1.2truein 
\BoxedEPSF{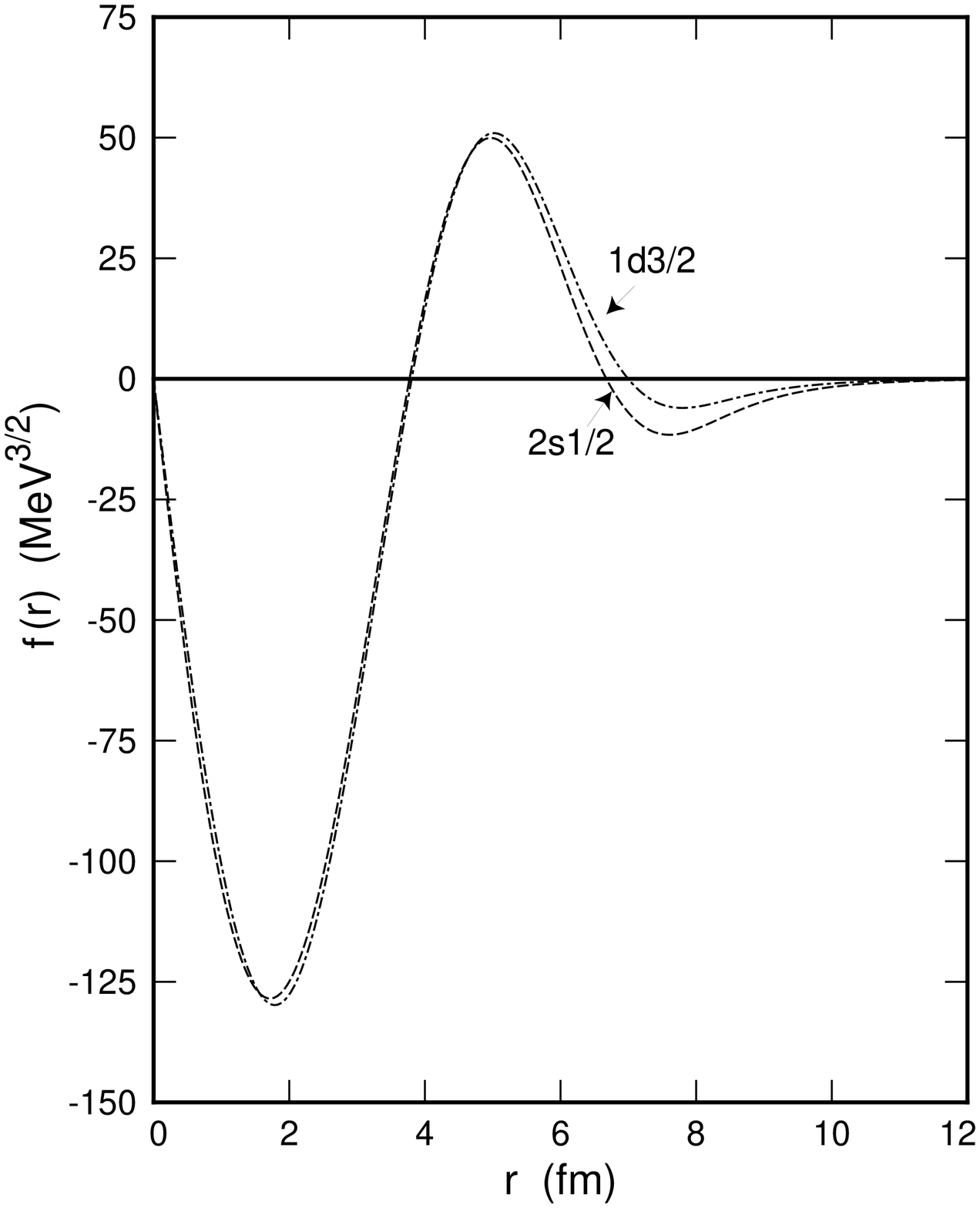 scaled 800}
\HideDisplacementBoxes
\hskip1.9truein 
\BoxedEPSF{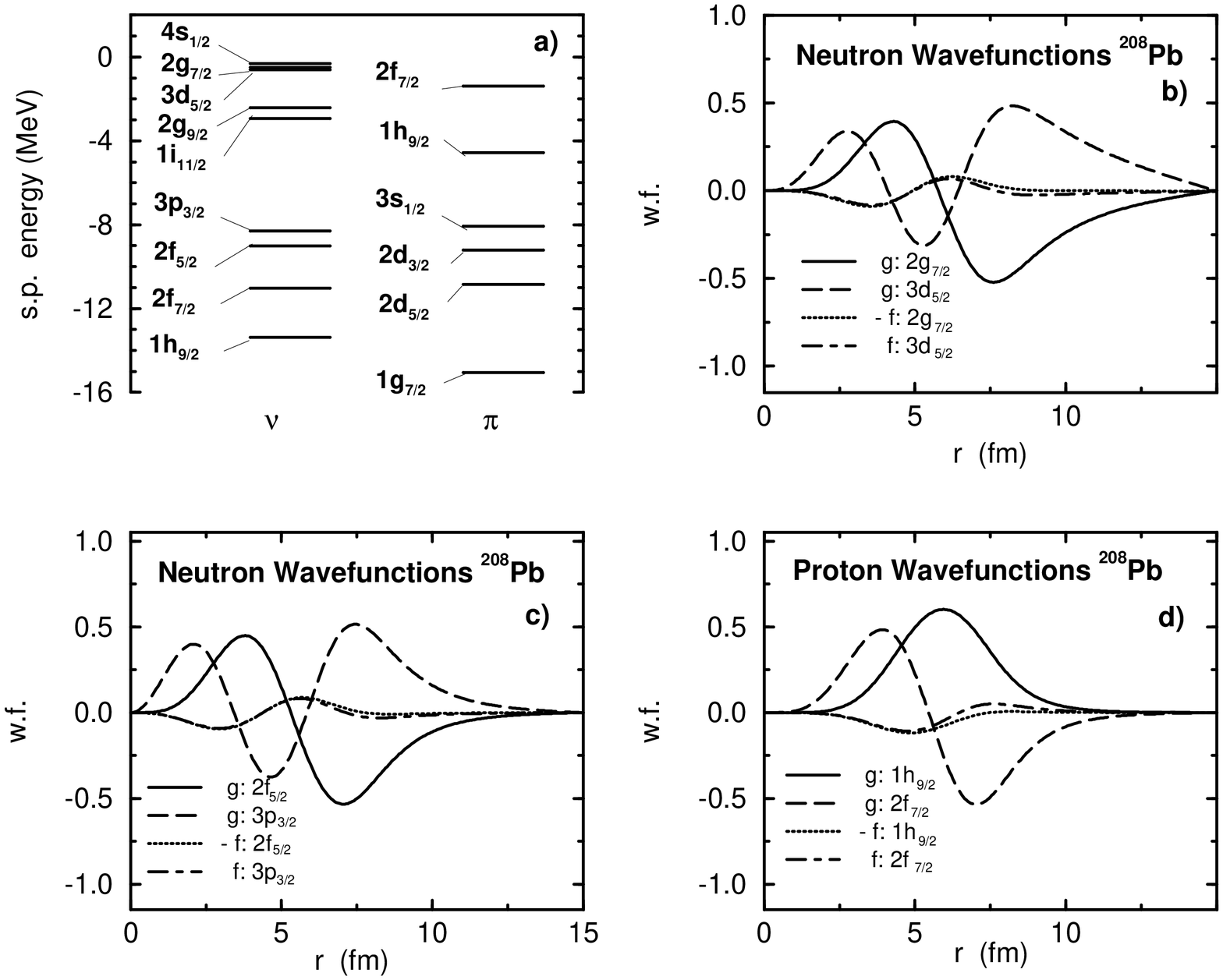 scaled 750}

\begin{thebibliography}{99}

\bibitem{dick}
Slansky R 1974 {\it Phys. Rep. } {\bf 11} 99

\bibitem{dick2}
Slansky R 1981 {\it Phys. Rep. } {\bf 79} 1

\bibitem{dick3}
Kass S, Moody R V, Patera J, and Slansky R 1990 {\it Affine Lie Algebras, Weight Multiplicities, and Branching Rules} (Berkeley:
University of California Press) 

\bibitem{kth}

Hecht K T and Adler A 1969 {\it Nucl. Phys.} {\bf A137} 129 

\bibitem{aa}

Arima A, Harvey M and Shimizu  K 1969 {\it Phys. Lett.} {\bf 30B} 517

\bibitem{gino}

Ginocchio J N 1997 {\it Phys. Rev. Lett.} {\bf 78} 436 

\bibitem{ami} 
Ginocchio J N  and  Leviatan A 1998 {\it Phys. Lett. B} {\bf 425} 1

\bibitem{bohr}

Bohr A, Hamamoto I and Mottelson  B R 1982 {\it Phys. Scr.} {\bf 26} 267

\bibitem {draayer3}
Beuschel T, Blokhin A L and Draayer J P 1997 {\it Nucl. Phys.} {\bf A619}
119
\bibitem{dudek}

Dudek J, Nazarewicz W, Szymanski Z and Leander G A 1987 {\it Phys. Rev.
Lett.} {\bf 59} 1405

\bibitem{twin}
Nazarewicz W, Twin P J, Fallon P and Garrett J D 1990 {\it Phys. Rev.
Lett.} {\bf64} 1654
\bibitem{stephens}
Stephens F S {\it et al} 1998 {\it Phys. Rev.} {\bf C57} R1565  
\bibitem{ben}

Mottelson B, 1991 {\it Nucl. Phys.} {\bf A522} 1  

\bibitem {draayer}

Blokhin A L, Bahri  C and Draayer J P 1995  {\it Phys. Rev. Lett.} {\bf
74} 4149


\bibitem{gino2}
Ginocchio J N  and  Madland D G 1998 {\it Phys. Rev. C} {\bf 57} 1167

\bibitem{mul}
Greiner W, M\"{u}ller  B and J. Rafelski 1985 {\it Quantum Electrodynamics of
Strong Fields} (New York: Springer-Verlag) 

\bibitem {bell}
Bell J S and  Ruegg H 1975 {\it Nucl. Phys.} {\bf B98} 151 

\bibitem{wal}
Serot B D and Walecka J D 1986 {\it The Relativistic Nuclear Many - Body
Problem} in {\it Advances in Nuclear Physics}, edited by J. W. Negele and
E. Vogt, Vol.\ {\bf 16} (New York: Plenum)

\bibitem{mad}

Nikolaus B A, Hoch  T and  Madland D G 1992 {\it Phys. Rev.} {\bf C46} 
1757

\bibitem{furn}

Cohen T D,  Furnstahl R J, Griegel K and  Jin X 1995  {\it Prog. in Part. and Nucl. Phys.} {\bf 35} 221

\bibitem {ring}
Lalazissis G A, Gambhir Y K, Maharana J P, Warke C S and Ring P 1998 {\it LANL archives} {\bf nucl-th/9806009} 

\bibitem {arima}
Meng J, Sugawara-Tanabe K, Yamaji S, Ring P and Arima A 1998 {\it Phys. Rev.} {\bf C58} R628 

\bibitem{cheng}
Cheng T. P and Li L. F. 1984 {\it Gauge Theory of Elementary Particle Physics} (New York: Oxford University Press) 

\bibitem {gino3}
Ginocchio J N, submitted to Phys Rev C; {\it LANL archives} {\bf nucl-th/9812025}.  

\end{thebibliography}
\end {document}